\documentclass{aa}
\usepackage{graphicx}
\usepackage{aalongtable}
\usepackage{deluxetable}
\usepackage{natbib}

\def\fermilat{\textit{Fermi}/LAT}
\def\fermi{\textit{Fermi}}

\bibpunct{(}{)}{;}{a}{}{,}
\begin{document}
\title{Radio--Optical--Gamma-Ray properties of MOJAVE AGN detected by \fermilat\ \thanks{Table~\ref{tab:1} is only available in electronic form at the CDS via
anonymous ftp to cdsarc.u-strasbg.fr (130.79.128.5).}}

\titlerunning{Multi-band properties of AGN detected by \fermilat\ }

\authorrunning{Arshakian et al.}

   \author{ T.G. Arshakian\inst{1,2}
          \and
          J. Le\'on-Tavares \inst{3}
          \and
          M. B\"ottcher  \inst{4}
          \and
          J. Torrealba \inst{5}
          \and
          V.H. Chavushyan \inst{5}
          \and
          M.L. Lister \inst{6}
          \and
          E. Ros \inst{7,1}
          \and
          J.A. Zensus \inst{1} 
          }

   \institute{ Max-Planck-Institut f\"ur Radioastronomie, Auf dem H\"ugel 69,
   53121 Bonn, Germany\\
   \email{tarshakian@mpifr-bonn.mpg.de}
          \and
    Byurakan Astrophysical Observatory, Byurakan 378433,
    Armenia and Isaac Newton Institute of Chile, Armenian
    Branch 
           \and     
    Aalto University Mets\"ahovi Radio Observatory, Mets\"ahovintie 114, 
    FIN-02540, Kylm\"al\"a, Finland\\
   \email{leon@kurp.hut.fi}
           \and
   Astrophysical Institute, Department of Physics and Astronomy, Ohio University,
   Athens, OH 45701, USA\\
   \email{boettchm@ohio.edu}
       \and
   Instituto Nacional de Astrof\'{\i}sica \'Optica y
   Electr\'onica, Apartado Postal 51 y 216, 72000 Puebla, Pue, M\'exico\\
   \email{[cjanet,vahram]@inaoep.mx}
          \and
   Department of Physics, Purdue University, 525 Northwestern Avenue,
   West Lafayette, IN 47907, USA \\
   \email{mlister@purdue.edu}
          \and
   Departament d'Astronomia i Astrof\'{\i}sica, Universitat de Val\`encia, E-46100 
   Burjassot, Spain \\
   \email{eduardo.ros@uv.es}
          }


   \date{Received <date> / Accepted <date>}


  \abstract
   {}
   {We use a sample of 83 core-dominated active galactic nuclei 
   (AGN) selected from the MOJAVE (Monitoring of Jets in AGN with  
   VLBA Experiments) 
   radio-flux-limited sample and detected with the \fermi\ Large Area Telescope (LAT) to 
   study the relations between non-simultaneous radio, optical, and $\gamma$-ray measurements.}  
   {We perform a multi-band statistical analysis to investigate the relations 
   between the emissions in different bands and reproduce these relations by 
   modeling of the spectral energy distributions of blazars.}
   {There is a significant correlation between the $\gamma$-ray luminosity 
   and the optical nuclear and radio (15~GHz) luminosities of blazars. We 
   report a well defined positive correlation between the $\gamma$-ray 
   luminosity and the radio-optical loudness for quasars and BL Lacertae type 
   objects (BL~Lacs). 
   A strong positive correlation is found between the radio  
   luminosity and the $\gamma$-ray-optical loudness for quasars, while a negative correlation 
   between the optical luminosity and the $\gamma$-ray-radio loudness is present for
   BL~Lacs. Modeling of these correlations with a simple leptonic jet model for blazars 
   indicates that variations of the accretion disk luminosity (and hence the jet 
   power) is able to reproduce the trends observed in most of the 
   correlations. To reproduce all observed correlations, variations of several parameters,
  such as the accretion power, jet viewing angle, Lorentz factor, and magnetic field of the  
  jet, are required.}
   {}

   \maketitle
%

\section{Introduction}

Blazars are observed over a wide range of the electromagnetic spectrum from radio to 
very-high-energy (VHE) $\gamma$-rays. Synchrotron emission, which is believed to
be dominant in the radio band, originates in the relativistic bipolar outflows 
which on milliarcsecond scales (parsec scales) appear as moving superluminal radio knots, which can be probed by very-long-baseline interferometry (VLBI) images \citep[e.g.,][]{ken98}. 

One quarter of all AGN of the radio-flux-limited MOJAVE \citep{lister09a}
sample objects were detected by the \fermi\ LAT during the first three months of operation 
\citep{lister09b}. For this sample of 31 blazars a significant correlation 
between radio (VLBA\footnote{Very Long Baseline Array}) flux density at 15~GHz and $\gamma$-ray photon flux was reported 
for quasi-simultaneous radio -- $\gamma$-ray measurements 
\citep{kovalev09}. They suggested that the radio and $\gamma$-ray emissions are produced in the 
cores of parsec-scale jets. A positive correlation between radio flux density at 8.4~GHz and $\gamma$-ray flux was reported by \cite{linford11} for 50 flat-spectrum quasars selected from the VLBA Imaging and Polarimetry Survey (VIPS). The correlation appeared to be stronger for radio bright quasars and the differences between
the $\gamma$-ray loud and quiet flat-spectrum quasars can be explained by Doppler boosting. Strong correlation between the $\gamma$-ray flux (above 100 MeV)
and the 20~GHz flux density was found for 134 flat-spectrum quasars and BL Lacs which were identified from 
cross correlation of the 1FGL with the 20~GHz Australia
Telescope Compact Array radio survey catalogue (Mahoni et al. 2010, Ghirlanda et al. 2011).
\cite{pushkarev10} studied 187 blazars selected 
from the First \fermi\ LAT catalog \citep[1FGL;][]{abdo10b} and found that 
on average the $\gamma$-ray flares precede the radio flares at 15~GHz with 
intrinsic time delays of 1.2 months. This result suggests that the origin of the bulk of the  the $\gamma$-ray photons is located within the 15~GHz VLBI core. High-frequency VLBI (43~GHz) and multi-waveband monitoring of blazars indicate that a single superluminal 
radio knot can cause a number of $\gamma$-ray flares, and that very rapid 
variability of $\gamma$-ray emission comes near the location of the 43~GHz 
cores of blazars jets \citep{marscher10}. Recently, \cite{tavares11} have shown that the strongest $\gamma$-ray flares follow the $mm$ flares in about 70 days. They interpret this as strong $\gamma$-ray events being produced at a distance of several parsecs downstream of the $mm$-wave core. 

The MOJAVE blazars show a tight connection between radio 
VLBA and optical emission. Long-term radio -- optical monitoring of radio galaxies showed 
that a variable optical emission on scales from a few months to a few years 
is generated in the innermost 0.4~pc region of the radio jet 
\citep{arshakian10a,tavares10}. A significant positive correlation at 
milliarcseconds scales was reported between the optical nuclear emission 
and (i) the radio emission of the unresolved cores at 15~GHz in superluminal 
quasars, and (ii) the extended jet emission in BL Lac objects by \cite{arshakian10b}. 
Those results suggested that the radio and optical emissions are beamed and 
originate in the innermost part of the sub-parsec-scale jet in quasars.

Extreme optical variability was a clear signature for some blazars
detected at $\gamma$-ray energies during the {\it CGRO}/EGRET era (Fichtel et
al. 1994). Taking advantage of the improved gamma sensitivity of \fermilat , 
recent multiwavelength variability studies on individual sources have
confirmed the tight connection among the $\gamma$-ray and variable 
optical emission \citep[e.g., 3C\,273 in][]{abdo10a}.

So far there was no attempt to compare optical continuum properties among
$\gamma$-ray blazars.  In this manuscript we study multi-band correlations 
to investigate the single production mechanism for continuum emission in 
different wavebands. Taking advantage of the unprecedented sensitivity 
provided by the 1FGL, we analyze relations between optical nuclear emission, 
radio emission on scales of milliarcseconds, and $\gamma$-ray emission of 
more than 80 MOJAVE AGN that had high confidence \fermilat\ source
associations after its first 11 months of operation. As discussed in
\cite{abdo10b}, it is possible that some
blazars may lack a \fermilat\ association due their location in a
particularly confused location (numerous $\gamma$-ray sources, or high
diffuse background near the galactic plane). Since the MOJAVE sample
includes only very radio-bright sources and avoids the galactic plane
region, we will refer for convenience in this paper to the non-LAT
associated MOJAVE sources as 'non-detected'. An interpretation of the 
correlations and the main physical process driving these correlations is discussed 
in the framework of a steady-state jet model.

The sample of blazars is defined in Sect.~2. In Sect.~3, the Kendall's 
$\tau$ statistical test is used to examine the correlations between 
emission characteristics defined in radio, optical, and $\gamma$-ray 
regimes. We interpret the implications of the jet model in Sect.~4 and 
summarize our results in Sect.~5. 

\begin{figure}
\center
  \includegraphics[width=8.5cm]{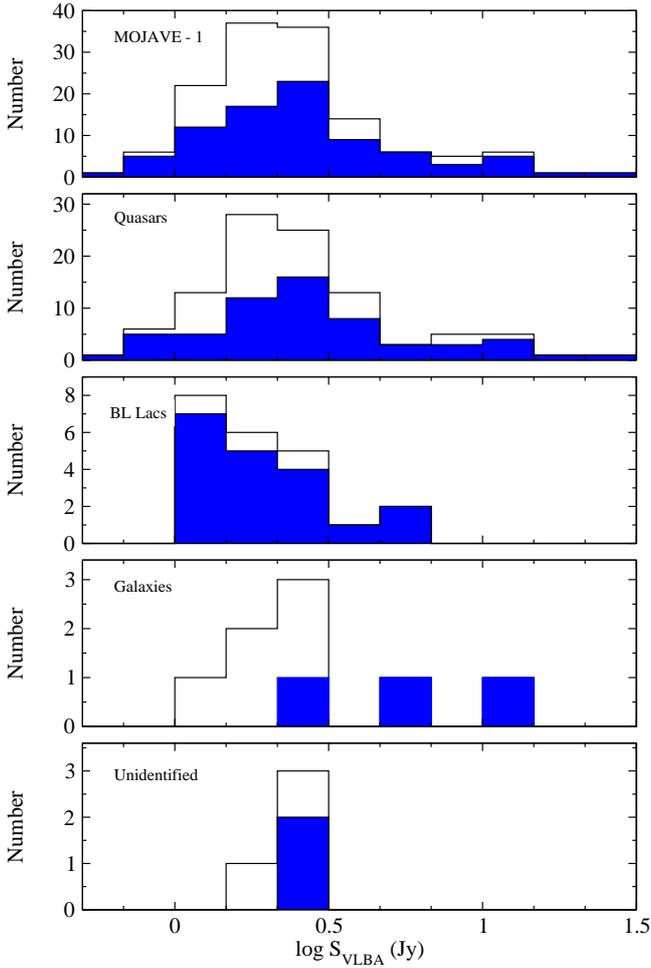}
 \caption{Distribution of radio VLBA flux densities at 15~GHz for 135 MOJAVE-1 AGN (full line) and AGN detected with \fermilat\  (filled blue histogram) for all blazars, quasars, BL Lacs, radio galaxies, and optically unidentified sources. }
   \label{fig:distofradioflux}
\end{figure}

Throughout the paper a flat cosmology with $H_{0} = 71$ km s$^{-1}$ Mpc$^{-1}$, 
$\Omega_{m} = 0.27$, and $\Omega_{\Lambda} = 0.73$ is adopted.

\begin{figure}
 \includegraphics[width=7.5cm]{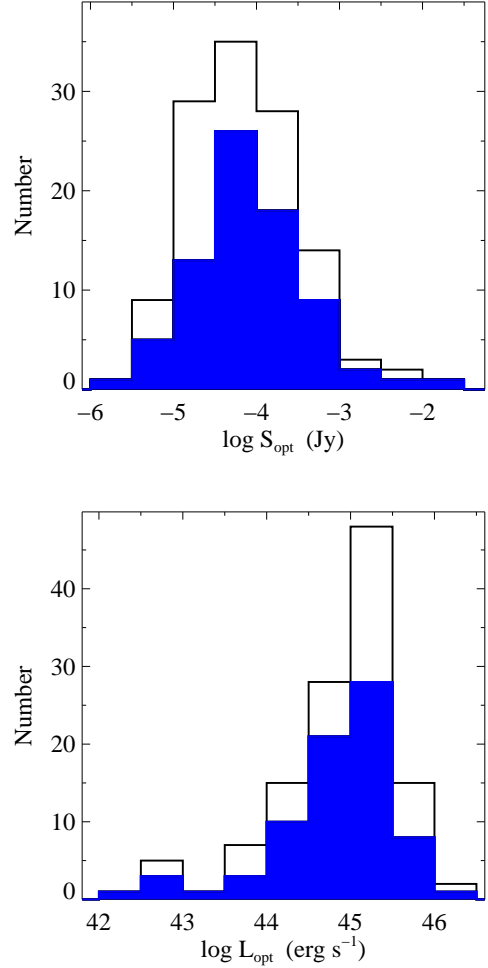}
 \caption{Distributions of optical nuclear fluxes (top panel) and luminosities (bottom panel) at 5100\,\AA\ for 122 AGN from the MOJAVE-1 sample (full line). Among these, 76 AGN are detected as the gamma-ray source (filled histogram) with \fermilat\ during the 1FGL period.}
  \label{fig:distofoptflux}
\end{figure}

\section{The sample of AGN detected with \fermilat}

We study a statistically complete sample of 135 core-dominated AGN compiled 
from the MOJAVE sample and the VLBA 15\,GHz 
monitoring survey \citep{ken98,zensus02,lister09a}. The complete sample, 
MOJAVE-1, is limited by the total VLBA (mas-scale) flux
density at 15\,GHz: it is greater 
than 1.5 Jy for sources in the Northern hemisphere ($\delta > 0$) and 
$> 2$ Jy in the Southern hemisphere ($-20 <\delta< 0$). The sample 
comprises quasars, BL Lacs, radio galaxies and some unidentified sources having typically a core-jet structure on milliarcsecond scales. At high redshifts, the sample has a selection bias towards BL Lacs and quasars with fast and bright one-sided jets oriented closer to the line of sight, while at low redshifts ($z<0.1$) there is a selection bias towards slow and intrinsically bright jets of radio galaxies (for more details see Lister \& Homan 2005, Arshakian et al. 2010b).  

The 1FGL catalogue includes 83 sources from the MOJAVE-1 sample (hereafter, 
M1-1FGL sample): 59 quasars, 19 BL~Lacs, 3 radio galaxies, and 2 unidentified 
sources. Most of the blazars not detected at $\gamma$-ray energies come from the weaker radio portion of the MOJAVE-1 sample (Fig.~\ref{fig:distofradioflux}) indicating that the $\gamma$-ray selected sample is biased towards radio strong blazars. The latter also  are strong $\gamma$-ray blazars since there is a strong
positive correlation between radio and $\gamma$-ray
fluxes of blazars (Kovalev et al. 2009).
 
For these 83 $\gamma$-ray selected sources, optical nuclear fluxes and redshift measurements were available for 76 objects \citep{arshakian10b}: 58 quasars, 15 BL~Lacs, and 3 radio galaxies (see Table~\ref{tab:1}). A procedure to correct the photometric fluxes for a stellar emission is discussed in \cite{arshakian10b}. Typical errors associated with optical nuclear fluxes are between $15\%$ and $25$\%.

We derive the optical nuclear luminosities ($L_{\rm opt}$), total VLBA 
luminosities ($L_{\rm VLBA}$), and the rest frame radio-optical loudness 
(the ratio of the $K$-corrected radio and optical fluxes, $R = S^{'}_{\rm VLBA}/S^{'}_{\rm opt}=S_{\rm VLBA} (1+z)^{-\alpha_{\rm r}+\alpha_{\rm o}}/ S_{\rm opt}
$, where $\alpha_{\rm r}=0$ is the spectral index of the total VLBA component,
$\alpha_{\rm o}$ is the optical spectral index adopted to be $-0.5$). We use the optical nuclear fluxes at 5100\,\AA\ given in \citet{arshakian10b} and the total radio flux densities 
at 15 GHz measured (during the period from 1994 to 2003 as part of the VLBA 15 GHz monitoring survey) at the epochs at which  
the unresolved component flux density has its maximum value. 
$\gamma$-ray luminosities were computed using 
the Eq.~(1) in \citet{ghisellini09}, 
\begin{equation}
L_{\gamma} = 4\pi d_{L}^2 \frac{S_{\gamma} (\nu_{1}, \nu_{2})}{(1+z)^{1-\alpha_{\gamma}}},
\end{equation}
where $S_{\gamma} (\nu_{1}, \nu_{2})$ 
is the energy flux between $\nu_{1}=0.1$~GeV and $\nu_{2}=100$~GeV from the 1FGL catalogue, $\alpha_{\gamma}$ is the
$\gamma$-ray energy index between the frequencies $\nu_{1}$ and $\nu_{2}$, and $d_L$ is the luminosity distance. Note 
that the $\gamma$-ray, optical, and radio luminosities of the MOJAVE-1 blazars are estimated from non-simultaneous observations and their typical uncertainties are between $\sim (10-15)$\% in $\gamma$-ray band, $\sim (15-25)$\% in optical, and $\la 5$\% \citep{kovalev05} in radio band. The rest frame $\gamma$-ray-radio loudness is defined as $G_r = S^{'}_{\gamma}/S^{'}_{\rm VLBA}=S_{\gamma}(1+z)^{-\alpha_{\gamma}+\alpha_{\rm r}}/S_{\rm VLBA}$ and the $\gamma$-ray-optical loudness is $G_o =S^{'}_{\gamma}/S^{'}_{\rm opt}= S_{\gamma}(1+z)^{-\alpha_{\gamma}+\alpha_{\rm o}}/S_{\rm opt}$.

We use the non-parametric Kendall's $\tau$ test to analyze correlations 
between independent variables and the partial Kendall's $\tau$ test \citep{akritas96} to 
account for the common dependence on redshift in the correlations between 
luminosities. Throughout the paper we assume a correlation to be significant 
if the chance probability for a null hypothesis is $P < 0.05$.

\section{Relations between $\gamma$-ray, optical, and radio properties of 
superluminal AGN}

\begin{figure}
  \includegraphics[width=\columnwidth]{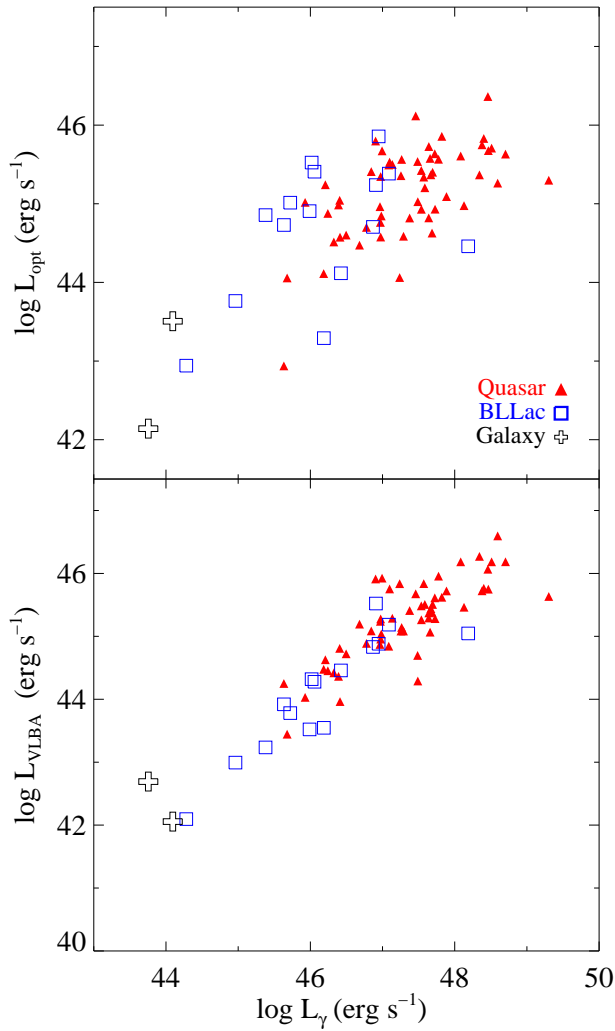}
 \caption{Optical nuclear luminosity against $\gamma$-ray luminosity (top panel), and radio VLBA (15~GHz) luminosity against $\gamma$-ray luminosity (bottom panel) for AGN from the M1-1FGL sample. Labels in the top panel denote the corresponding population.
}
   \label{fig:lum-lum}
\end{figure}

\emph{Detection rate of $\gamma$-ray blazars. }
The relative number of blazars detected in $\gamma$-rays increases with
increasing radio flux, and most quasars not detected in $\gamma$-rays are
radio weak (Fig.~\ref{fig:distofradioflux}).
This is explained by a tight correlation between radio and $\gamma$-ray emission of quasars (Kovalev et al. 2009). Most of BL Lacs and radio bright galaxies are detected in $\gamma$-ray band.

The distributions of optical nuclear fluxes for detected and not detected $\gamma$-ray blazars have similar ranges (Fig.~\ref{fig:distofoptflux}; top panel). The Kolmogorov-Smirnov (K-S) statistical test shows that the null hypothesis that these distributions are drawn from the same parent population can not be rejected ($P=0.15$; confidence 
level of $85$\,\%). We conclude that among radio-selected blazars, the detection rate of $\gamma$-ray AGN does not depend on their optical nuclear fluxes, and most of undetected $\gamma$-ray blazars are in fact quite optically luminous (Fig.~\ref{fig:distofoptflux}; bottom panel).

\begin{figure}
 \includegraphics[width=0.53\textwidth]{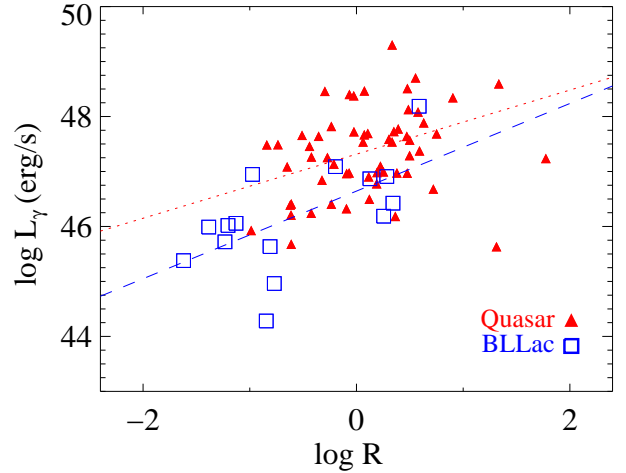}
 \caption{$\gamma$-ray luminosity ($L_{\gamma}$) versus radio-optical loudness ($R\equiv S^{'}_{\rm VLBA}/S^{'}_{\rm opt}$) for the M1-1FGL quasars and BL Lacs. The dotted and dashed lines represent the ordinary least-square fit to the data for quasars and BL Lacs respectively.}
 \label{fig:glum-rl}
\end{figure}

\begin{table*}[ht]
  \caption[]{Kendall's $\tau$ correlation analysis between emission characteristics of AGN from the M1-1FGL sample. A1 and A2 are the independent variables for which the Kendalls $\tau$ correlation 
analysis is performed, and A3 is the control variable (if
exists then the partial KendallÕs $\tau$ correlation analysis is applied to A1 and A2), 
 $\tau$ is the correlation coefficient, and $P$ is the 
probability of a chance correlation. The correlations are considered to be significant if the chance probability $P < 0.05$ (or confidence level $>95$\,\%).}
\label{table:statistics}
\begin{center}
\begin{tabular}{cccrccrccrccccc}
 \hline \hline
   &  &  & \multicolumn{2}{c}{All} &  &  \multicolumn{2}{c}{Quasars} & & \multicolumn{2}{c}{BL Lacs}\\
 \cline{4-5}\cline{7-8}\cline{10-11} \smallskip
  A1 & A2 & A3 & $\tau$ & $P$ &  &  $\tau$ & $P$ & & $\tau$ & $P$ \\
 \hline
$ L_{\rm VLBA}$ & $ L_{\rm opt}$  &  $z$ & 0.22 & 0.004 & & 0.25 & 0.006 & &0.01 & 0.957\\
$ L_{\rm \gamma}$ & $ L_{\rm opt}$ & $z$ &  0.19 & 0.015 & & 0.23 & 0.011 & &0.03 & 0.899\\
$ L_{\rm \gamma}$ & $ L_{\rm VLBA}$ &  $z$ & 0.41 & $2\times 10^{-7}$& &0.39 & $2\times 10^{-5}$ & &0.42 & 0.036\\
$ L_{\rm \gamma}$ &  $S^{'}_{\rm VLBA}/S^{'}_{\rm opt}$& - & 0.36 & $4\times 10^{-6}$& &0.24 & 0.008 & &0.43 & 0.025\\
$ L_{\rm VLBA}$   & $S^{'}_{\gamma}/S^{'}_{\rm opt}$ & - & 0.41 & $2\times 10^{-7}$ & & 0.34 & 0.0002 & & 0.16 & 0.4\\
$ L_{\rm opt}$   & $S^{'}_{\gamma}/S^{'}_{\rm VLBA}$ & - & $-$0.11 &  0.11 & &0.16 & 0.071 & &$-$0.39 & 0.042\\
 \hline
\end{tabular}
\end{center}
\end{table*}

\emph{Optical -- Radio emission.} \citet{arshakian10b} found a positive correlation 
in the $L_{\rm opt}$ -- $L_{\rm VLBA}$ plane for AGN from the MOJAVE-1 
sample. They concluded that the correlation is due to the population of quasars 
and that the optical emission is non-thermal and generated in the parsec-scale jet.
This is supported by studies of individual radio galaxies, 3C\,390.3 and 3C\,120 
\citep{arshakian10a,tavares10} for which a link between optical continuum 
variability and kinematics of the parsec-scale jet was found. It was interpreted 
in terms of optical continuum flares generated at subparsec-scales in the innermost 
part of a relativistic jet rather than in the accretion disk. 

\emph{Gamma-ray -- Optical emission.} No correlation is found for BL Lacs, while a 
significant positive correlation between $L_{\gamma}$ and $L_{\rm opt}$ is found 
for quasars (99\,\%; Fig.~\ref{fig:lum-lum}; Table~\ref{table:statistics}) 
suggesting a single production mechanism for $\gamma$-ray and optical nuclear 
emission.

\emph{Gamma-ray -- Radio emission.} \citet{kovalev09} reported a correlation 
between $\gamma$-ray and radio VLBA (8~GHz) emission for a sample of $\sim 30$
AGN. This correlation holds at high confidence level ($>99.9$\,\%; 
Fig.~\ref{fig:lum-lum}) for non-simultaneous measurements at 15~GHz and a larger 
sample (M1-1FGL) of AGN. This suggests that the powers averaged over the 
long time scales are correlated and, hence, the Doppler-factors of the 
parsec-scale jet in the gamma and radio domains are not changing substantially 
on a timescale of a few years. \citet{pushkarev10} found that the $\gamma$-ray
emission leads the radio emission of the parsec-scale jet at 15~GHz with a time 
delay of a few months. They interpreted the observed time lag as a result of 
synchrotron opacity in the jet: the radio and $\gamma$-ray emissions can be  
generated in the same region by perturbations of the jet \citep{marscher85} and become observable 
with some time delay due to the opacity effects. If this scenario is correct, 
the variable optical emission is also generated in the perturbation moving 
upstream in the jet, and one should expect that the optical emission leads the 
radio emission and is also delayed with respect to the $\gamma$-ray emission. 

The $L_{\rm VLBA}$ -- $L_{\gamma}$ correlation is visibly tighter (0.39 dex scatter) than 
the $L_{\rm opt}$ -- $L_{\gamma}$ one (0.54 dex scatter; Fig.~\ref{fig:lum-lum}).
The remarkably tight relation between $L_{\rm VLBA}$ and $L_{\gamma}$, $\log L_{\rm VLBA}=(0.83\pm0.03) \log L_{\gamma} + (5.96\pm0.26)$, over a range of about five orders of magnitudes, can be used to estimate the $\gamma$-ray flux of not detected $\gamma$-ray blazars from their VLBA flux density.
One may think that the corrected optical nuclear emission of some AGN 
\citep{arshakian10b} might be contaminated by a contribution of a stellar
component, thus causing the large dispersion in the $L_{\rm opt}$ -- $L_{\gamma}$ 
diagram. This contamination should be stronger in radio galaxies and weaker 
in quasars and BL~Lacs for which the contribution of optical nuclear emission 
is dominant. The large dispersion in the $L_{\rm opt}$ -- $L_{\gamma}$ plane 
can be due to 
stronger variability in the optical regime than in the radio, and/or a wider range of 
Doppler factors in the optical regime compared to the range of Doppler factors 
of the jet at the location where the 15~GHz emission is produced, if the bulk 
of the optical emission is generated in the relativistic jet and it is Doppler 
boosted. 

We report a significant positive correlation ($> 99\,\%$) between $L_{\gamma}$ 
and radio-optical loudness $R$ for quasars and BL Lacs (see Fig.~\ref{fig:glum-rl} and 
Table~\ref{table:statistics}). For the combined sample the $L_{\gamma} \propto R^{0.95\pm 0.11}$, for quasars $L_{\gamma} \propto R^{0.44\pm 0.18}$, and for BL~Lacs 
$L_{\gamma} \propto R^{0.79\pm 0.27}$. The slope derived 
for quasars is shallower than for BL~Lacs. 
To test the robustness of this result, we excluded one outlier quasar ($\log R=1.31$ and $\log L_{\gamma}=45.63$) and repeated the fitting of data. The slope for quasars increases, $0.58 \pm 0.17$, and coincides with the slope of BL Lac objects within the error range. We conclude that the slopes for quasars and BL Lacs are not statistically different. A noticeable difference in radio-optical loudness is found between populations of quasars and BL~Lacs: $\log R=0.12$ for quasars and $\log R=-0.81$ for BL~Lacs.

\begin{figure*}
 \includegraphics[width=\textwidth]{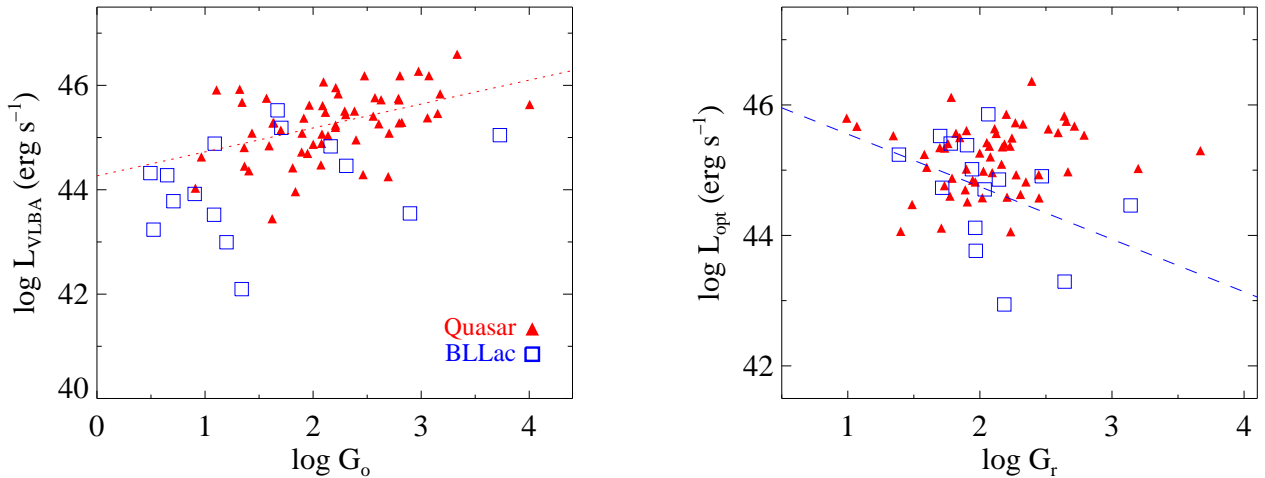}
 \caption{\emph{Left panel.} Radio VLBA luminosity against $\gamma$-optical loudness ($G_o \equiv S^{'}_{\gamma}/S^{'}_{\rm opt}$) for the M1-1FGL quasars and BL~Lacs. \emph{Right panel.} Optical luminosity against  $\gamma$-radio loudness  ($G_r \equiv S^{'}_{\gamma}/S^{'}_{\rm VLBA}$). The dotted and dashed lines represent the ordinary least-square fit to the data for quasars and BL Lacs respectively.}
   \label{fig:rlum-gl}
\end{figure*}

\emph{Gamma-ray loudness.} We define the rest frame $\gamma$-optical loudness 
($G_{o}$) and $\gamma$-radio loudness ($G_{r}$) as $G_o \equiv S^{'}_{\gamma}/S^{'}_{\rm opt}$ 
and $G_r \equiv S^{'}_{\gamma}/S^{'}_{\rm VLBA}$, respectively. We find that the radio 
luminosity and  $\gamma$-optical loudness are significantly correlated only for quasars (see Table~\ref{table:statistics} and Fig.~\ref{fig:rlum-gl}, left panel), $L_{\rm VLBA} \propto G_{o}^{0.45 \pm 0.12}$. 

The correlation in the $L_{\rm opt}$ -- $G_{r}$ plane (Fig.~\ref{fig:rlum-gl}, right panel) is marginally significant only for BL Lacs (Table~\ref{table:statistics}). A large amount of scatter about the fitted line and small sampling of BL Lacs account for the large error of the power index ($L_{\rm opt} \propto G_r^{-0.81\pm 0.48}$).

Logarithm of the median $\gamma$-optical loudness is significantly different for quasars and BL~Lacs ($2.1$ and $1.2$), while there is no difference between their medians of $\gamma$-radio loudness (2.1 and 2). 
There are no high-energy peaked BL~Lacs in the MOJAVE-1 sample. The latter have
significantly different SED properties than the radio-selected
MOJAVE-1 BL Lacs (Lister et al. 2011). The differences found between quasars and BL~Lacs are also supported 
in recent studies (e.g., Ghisellini et al. 2009, Sambruna et al. 2010, Tornikoski 
et al. 2010, Gupta et al. 2011) suggesting the presence of different physical conditions along 
the jet in quasars and BL~Lacs. This and other insights into the relationship 
between the $\gamma$-ray, radio, and optical emission will be pursued in more 
detail in further studies. \\

Relations found between $L_{\gamma}$ and $R$, $L_{\rm VLBA}$ and $G_{o}$, 
and $L_{\rm opt}$ and $G_{r}$ are due to variations of the SED of blazars, 
which can be driven by changes of the Doppler factor, magnetic field, and 
intrinsic power of the jets. In the next section, we attempt to identify 
the main physical parameters of the jet responsible for the observed 
correlations.

\section{Interpretation and model implications}

The broadband emission of blazars is generally understood as originating
from ultrarelativistic particles in the relativistic jet, which is oriented
at a small angle $\theta_{\rm obs} \sim 1/\Gamma$ (where $\Gamma$ is the
bulk Lorentz factor of the material moving along the jet) with respect to
our line of sight. The low-frequency continuum emission from radio to
optical -- UV (or even X-rays, in high-frequency-peaked BL~Lac objects)
frequencies is generally accepted to be synchrotron emission from relativistic
electrons in the jet. The high-energy (X-ray--$\gamma$-ray) emission
can be either inverse Compton emission from the same electrons (leptonic models),
or emission resulting from the production of secondaries in photo-pair
and photo-pion production reactions of ultrarelativistic hadrons in
the jet (hadronic models). For a recent review of blazar models, see,
e.g., \cite{boettcher10}. 

\begin{figure}
    \includegraphics[width=\columnwidth]{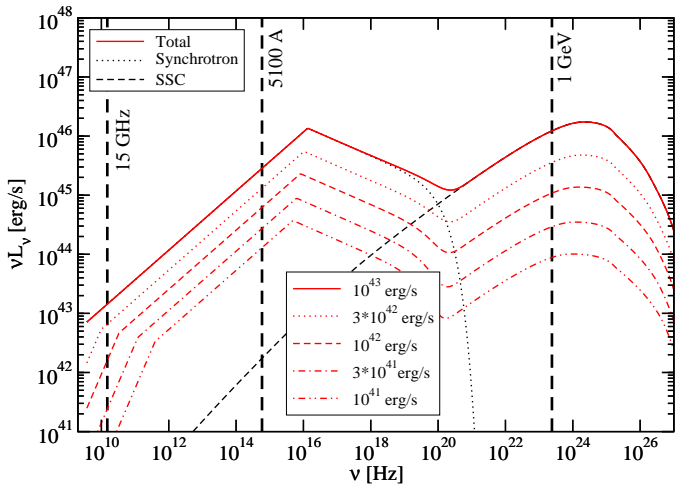}
\caption{Sequence of model SEDs for BL Lac objects, varying the jet power, 
while adjusting the magnetic field to equipartition with the relativistic 
electrons in the jet. The curves are labelled by the value of $L_j$. 
The black dotted and dashed lines indicate the 
individual radiation components adding up to the the total spectrum with 
the highest $L_j = 10^{43}$~erg~s$^{-1}$. This sequence yields the points 
shown as blue open squares in Figs. \ref{RLgamma} -- \ref{SgoLradio} . }
\label{BLLacSEDs}
\end{figure}

We will discuss a possible interpretation of the correlations described above, in the framework of a steady-state, one-zone leptonic model. Specifically, we adopt the semi-analytical description of a generic leptonic blazar model of \cite{bd02}. In
this model, a power-law distribution of relativistic electrons,
$Q(\gamma) = Q_0 \, \gamma^{-q} \, H(\gamma; \gamma_1, \gamma_2)$ 
with low- and high-energy cut-offs $\gamma_1$ and $\gamma_2$ is
persistently injected into a spherical emission region of radius
$R_B$, which moves with constant speed $\beta_{\Gamma}\,c$, corresponding
to bulk Lorentz factor $\Gamma$, along the jet. In order to reduce the
free parameter space, we choose an observing angle of $\theta_{\rm obs}
= 1/\Gamma$ so that the Doppler factor $D = \left( \Gamma \, [1
- \beta_{\Gamma} \cos\theta_{\rm obs} ] \right)^{-1} = \Gamma$.
A tangled magnetic field $B$ is present in the emission region. 
Modeling of the SEDs of many BL~Lac objects indicates that a model
including only synchrotron and synchrotron-self-Compton (SSC), is
often sufficient to describe those SEDs 
\citep[e.g.,][]{mk97,pian98,petry00,krawczynski02}. 
Therefore, we adopt a pure SSC model in this study to represent the 
SEDs of high-frequency-peaked blazars (BL~Lac objects), which is
fully determined through the model parameters listed above.
In the model of \cite{bd02}, the injected electron distribution
is followed in an iterative process to equilibrium between ongoing
particle injection and radiative cooling and escape. Fig.~\ref{BLLacSEDs}
shows a characteristic sequence of BL Lac SEDs resulting from a variation
of the jet power $L_j$, while keeping the magnetic field at equipartition
with the equilibrium electron distribution. Typical parameters for one
of those SED calculations are listed in Table~\ref{bllacparameters}.

\begin{figure}
   \includegraphics[width=\columnwidth]{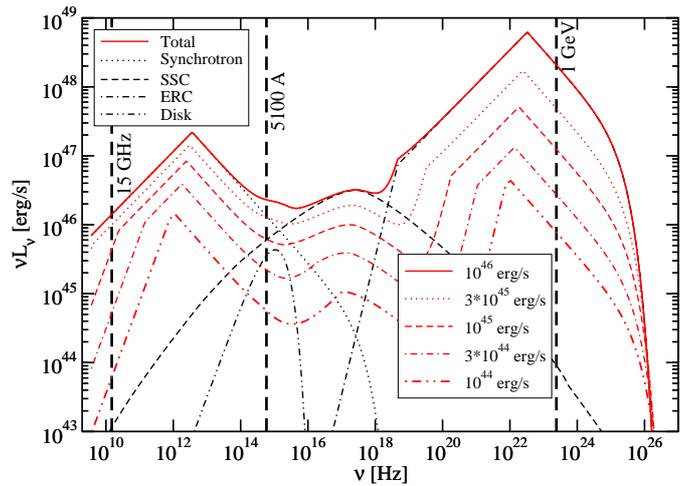}
\caption{Sequence of model SEDs for quasars, varying the accretion disk luminosity
$L_D$ (and hence, the jet power $L_j = f_j L_D$), while adjusting the magnetic
field to equipartition with the relativistic electrons in the jet. The curves
are labelled by the value of $L_D$. The black dotted, dashed, dot-dashed, and
dot-dot-dashed lines indicate the individual radiation components adding up to the total spectrum with the highest $L_D = 10^{46}$~erg~s$^{-1}$. This sequence yields the points shown as blue filled squares in Figs.~\ref{RLgamma} -- \ref{SgoLradio}. }
\label{quasarSEDs}
\end{figure}

For low-frequency-peaked BL Lacs (LBLs) and flat-spectrum radio quasars (FSRQs), 
it is generally accepted that leptonic models require external photon sources 
for Compton scattering (EC = External Compton) to produce the $\gamma$-ray 
emission \citep[e.g.,][]{dss97,sambruna97,muk99,madejski99,bb00,hartman01}. 
In the case of quasars, the most prominent external radiation field might
be the quasi-isotropic radiation field of the broad-line region (BLR), as 
long as the emission region is located within the BLR. We model the BLR
as a spherical shell at a distance $R_{\rm BLR}$ from the central engine,
reprocessing a fraction $\tau_{\rm BLR}$ of the accretion disk luminosity,
$L_D$. In this case, the jet power is parameterized as a fraction $f_j$ 
of the accretion disk luminosity, i.e., $L_j = f_j \, L_D$. As in the
case of BL~Lac objects, we assume the persistent injection of a power-law 
distribution of ultrarelativistic electrons into the emission region, and
let the distribution evolve to an equilibrium between injection, radiative
cooling, and escape. A typical set of parameters producing an FSRQ-like
SED is listed in Table~\ref{quasarparameters}. A sequence of FSRQ SEDs, 
resulting from a change in the accretion disk power (and hence also the 
jet power), while adjusting the magnetic field to be in equipartition with 
the equilibrium electron distribution, is shown in Fig.~\ref{quasarSEDs}.

\begin{figure*}
\centering
  \includegraphics[width=0.75\textwidth]{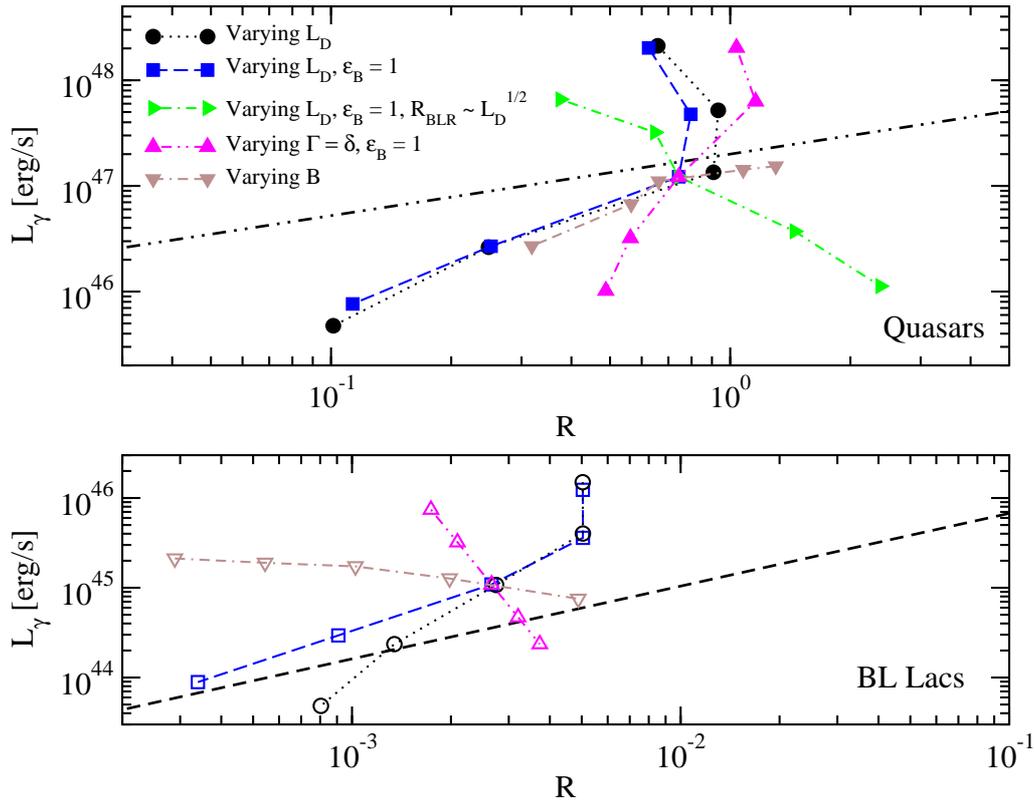}
\caption{Model correlations of radio-optical loudness $R$ vs. gamma-ray luminosity
$L_{\gamma}$ along sequences varying individual parameters. Each sequence represents the effect
of a variation of one individual base parameter. The legend in the
top left indicates the parameters varied along the sequences. Filled symbols
indicate SEDs characteristic of FSRQs, while open symbols indicate sequences
of SEDs characteristic of BL Lacs. The long-dashed and dot-dot-dashed lines
mark the power-law correlations that best fit the observational data. The 
sequences marked in blue dashed line (squares), showing reasonable agreement with the best-fit
correlations, correspond to the sequences of SEDs shown in Figs.~\ref{BLLacSEDs} and \ref{quasarSEDs} (see also Tables~\ref{bllacparameters} and \ref{quasarparameters}). }
\label{RLgamma}
\end{figure*}

\begin{figure*}
\centering
  \includegraphics[width=0.75\textwidth]{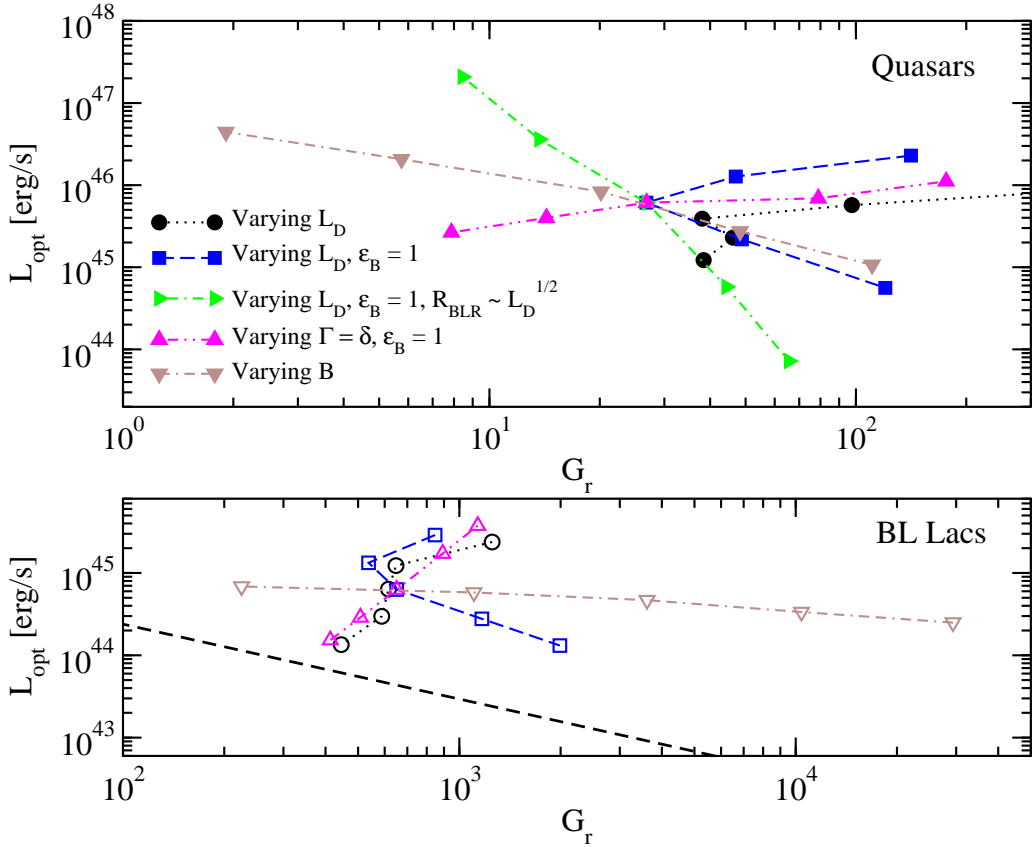}
\caption{Model correlations of the $\gamma$-ray-to-radio flux ratio vs. optical 
luminosity along sequences varying individual parameters. 
Each sequence represents the effect
of a variation of one individual base parameter.
The legend in the
top left indicates the parameters varied along the sequences. Filled symbols
indicate SEDs characteristic of FSRQs, while open symbols indicate sequences
of SEDs characteristic of BL Lacs. The long-dashed line
marks the power-law correlation that best fits the observational data. The 
sequences marked in blue dashed line (squares) correspond to the sequences of SEDs shown in Figs.~\ref{BLLacSEDs} and \ref{quasarSEDs} (see also Tables~\ref{bllacparameters} and \ref{quasarparameters}). }
\label{SgrLopt}
\end{figure*}

\begin{figure*}
\centering
  \includegraphics[width=0.75\textwidth]{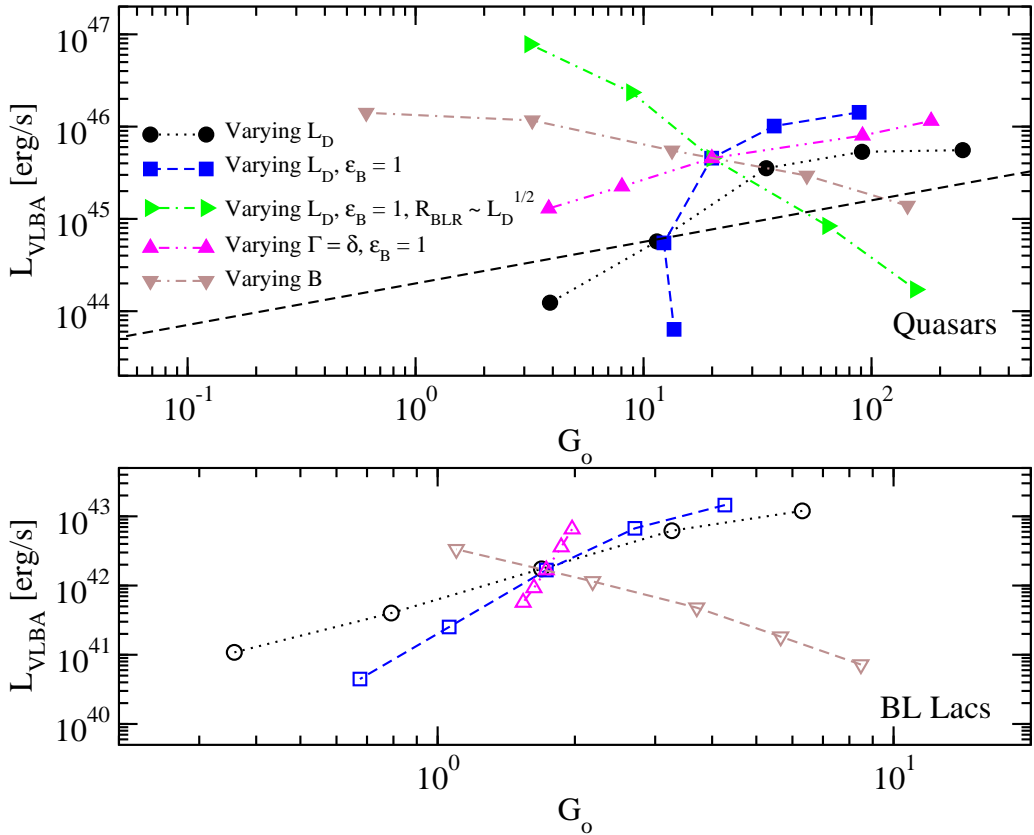} 
\caption{Model correlations of the $\gamma$-ray-to-optical flux ratio vs. radio 
luminosity along sequences varying $L_D$ (for quasars, with $f_j = const.$) and 
$L_j$ (for BL Lacs), keeping the magnetic field at equipartition (squares), and 
varying $B$ (downward triangles). Filled symbols
indicate SEDs characteristic of FSRQs, while open symbols indicate sequences
of SEDs characteristic of BL Lacs. The 
sequences marked in blue dashed line (squares) correspond to 
the sequences of SEDs shown in Figs.~\ref{BLLacSEDs} and \ref{quasarSEDs}. }
\label{SgoLradio}
\end{figure*}

We tested whether variations of individual parameters for FSRQs and BL~Lacs
were able to reproduce the observed statistically significant ($P<0.05$) correlations between $R$ and $L_{\gamma}$, $G_{o}$ and $L_{\rm VLBA}$ for quasars, 
and between $G_{r}$ and $L_{\rm opt}$ for BL Lacs. In particular, we tested whether these correlations
can arise from a change in the following fundamental parameters: (a) Accretion
disk (and hence, jet) power $L_D$ ($L_j$), (b) Accretion disk (and jet) power,
while keeping the magnetic field in equipartition with the equilibrium electron
distribution ($\epsilon_B \equiv u_B/u_e = 1$), (c) Accretion disk (and jet) 
power with $\epsilon_B = 1$, adjusting the BLR radius as $R_{\rm BLR} \propto
L_D^{1/2}$, (d) Lorentz factor $\Gamma$ (and, hence, the Doppler factor $D$),
(e) viewing angle $\theta_{\rm obs}$ (also changing the Doppler factor $D$),
and (f) magnetic field $B$. The resulting theoretical correlations for FSRQs are
shown by the filled symbols in Figs.~\ref{RLgamma} -- \ref{SgoLradio}, 
while open symbols show representative sequences for BL~Lac objects. In all cases, we ran one sequence per base parameter. Figs.~\ref{RLgamma} -- \ref{SgoLradio} show representative results. The straight lines in these figures indicate the best
power-law fits to the observed correlations. 

\begin{table}
\caption{Typical leptonic jet model parameters for a BL~Lac object.}             
\label{bllacparameters}      
\centering                          
{\tiny
\begin{tabular}{c c c}        
\hline\hline                 
Quantity & Symbol & Value  \\    
\hline                        
Minimum injection electron Lorentz factor & $\gamma_1$ & $10^4$ \\
Maximum injection electron Lorentz factor & $\gamma_2$ & $10^6$ \\
Electron injection spectral index & q & 2.5 \\
Magnetic field & $B$ & 2~G \\
Blob radius & $R_B$ & $3 \times 10^{15}$~cm \\
Bulk Lorentz factor & $\Gamma$ & 15 \\
Observing angle & $\theta_{\rm obs}$ & $3.82^o$ \\
Jet power & $L_j$ & $10^{42}$~erg~s$^{-1}$ \\
\hline                                   
\end{tabular}
}
\end{table}

\begin{table}
\caption{Typical leptonic jet model parameters for a quasar.}             
\label{quasarparameters}      
\centering                          
{\tiny
\begin{tabular}{c c c}        
\hline\hline                 
Quantity & Symbol & Value  \\    
\hline                        
Minimum injection electron Lorentz factor & $\gamma_1$ & 300 \\
Maximum injection electron Lorentz factor & $\gamma_2$ & $10^5$ \\
Electron injection spectral index & q & 3.1 \\
Magnetic field & $B$ & 0.661~G \\
Blob radius & $R_B$ & $10^{17}$~cm \\
Bulk Lorentz factor & $\Gamma$ & 15 \\
Observing angle & $\theta_{\rm obs}$ & $3.82^o$ \\
Accretion disk luminosity & $L_D$ & $10^{45}$ erg s$^{-1}$ \\
Jet power fraction & $f_j$ & 0.05 \\
Reprocessing Optical Depth of BLR & $\tau_{\rm BLR}$ & 0.1 \\
Effective radius of BLR & $R_{\rm BLR}$ & $10^{18}$~cm \\
\hline                                   
\end{tabular}
}
\end{table}

When comparing the theoretical correlations with the observed ones, one needs
to keep in mind that our model only takes into account radio emission from the
high-energy emission region in the sub-pc scale core. Realistically, we expect 
a substantial contribution to the observed radio emission to arise on larger 
($\ga$~pc) scales. Therefore, our model radio fluxes should be considered
as lower limits. In particular, additional radio emission from larger scales
would shift all model curves to the right in Fig.~\ref{RLgamma} and to the
left in Fig.~\ref{SgrLopt}. Considering this potential offset, we identified
a set of parameters with which the observed $R$ vs. $L_{\gamma}$ (Fig.~\ref{RLgamma}), $G_{r}$ vs. $L_{\rm opt}$ (Fig.~\ref{SgrLopt}), and $G_{o}$ vs. $L_{\rm VLBA}$ (Fig.~\ref{SgoLradio})  
correlations can be reasonably well reproduced through a variation of 
$L_D$ (and $L_j$), while keeping the magnetic field in equipartition 
with the electron distribution. Those sequences are marked by the blue dashed line (square symbols) in Figs.~\ref{RLgamma} -- \ref{SgoLradio}, and the sequences 
of SEDs are illustrated in Figs.~\ref{BLLacSEDs} and \ref{quasarSEDs} for typical 
parameters of the jet in BL Lacs and quasars (Tables~\ref{bllacparameters} and \ref{quasarparameters}). Most other parameter sequences produce correlations with 
opposite slopes than observed, while a variation of only the magnetic field over about 
two orders of magnitude results in too small a range in radio-optical loudness 
to be consistent with the observed correlations (downward triangles in 
Fig.~\ref{RLgamma}). 

However, while reasonable agreement can be achieved with the $R$ vs. 
$L_{\gamma}$ correlation for quasars and BL Lacs and $G_{r}$ vs. $L_{\rm opt}$ 
correlation for BL Lacs, our favored sequence of changing $L_D$ ($L_j$) with 
$\epsilon_B \equiv 1$ predict steep-slope positive correlations
between $G_{o}$ and $L_{\rm VLBA}$ for quasars, in discrepancy with the slope of the observed positive correlation (Fig.~\ref{fig:rlum-gl}; left panel). A better agreement is achieved for changes of the accretion power $L_D$ and the Lorentz factor of the jet (black circles and downward triangles in Fig.\ref{SgoLradio}).
We therefore conclude that a simple 
one-parameter sequence is not able to reproduce the observational 
results. The substantial scatter in all correlations also might be an indication that 
those are not single-parameter sequences. In a realistic scenario, we expect that a combination of different parameter variations and physical relations between parameters are responsible for the observed range 
of radio, optical, and $\gamma$-ray luminosities. To disentangle the effect of multiple parameter sequences it is essential to provide information on individual parameters (e.g. variance in Lorentz factor, variance in $B$ field, etc.) from further studies.
Moreover, the dispersion of the observed correlations can be used to constrain 
the distribution of initial parameters of the model. For this 
purpose, (quasi-) simultaneous observations of the SED (from radio to $\gamma$-ray) of more than 100 blazars \citep{aatrokoski11,giommi11} may be used to test the found correlations and to constrain further the jet model. This will be presented in a subsequent paper.

Predictions of the model sequences are presented for quasars in Fig.~\ref{SgrLopt} (top panel) and BL~Lacs in Fig.~\ref{SgoLradio} (empty symbols). These can be tested with significant $L_{\rm opt}$ -- $G_{r}$ and  $L_{\rm VLBA}$ -- $G_{o}$ correlations derived from larger samples of blazars.

\section{Summary}
Using the sample of 78 core dominated blazars (59 quasars and 19 BL~Lacs) 
detected by \fermilat\ from the MOJAVE-1 sample, we investigate the relations between their radio, 
optical, and $\gamma$-ray emission available from non-simultaneous 
observations. Our main results are summarized as follows:

\begin{itemize}
\item The detection rate of the $\gamma$-ray blazars from the radio-selected MOJAVE-1 sample is high for radio bright blazars and it is independent of their optical nuclear fluxes. 
Most of non-detected $\gamma$-ray blazars are radio weak quasars. They are optically faint but also very luminous, $L_{\rm opt}>10^{44.5}$ erg s$^{-1}$.

\item 
The known correlation between $\gamma$-ray flux and compact radio flux density 
(measured quasi-simultaneously) is also valid for non-simultaneous  
measurements of $L_{\gamma}$ and $L_{\rm VLBA}$ for 59 quasars and 
19 BL~Lacs. This indicates that the Doppler-factors of the jet-plasma in the gamma and radio regimes are fairly constant on a timescale of a few years. The $L_{\gamma}$ -- $L_{\rm VLBA}$ correlation is significantly 
stronger than that in the $L_{\gamma}$ -- $L_{\rm opt}$ plane. The latter 
holds exclusively for quasars at the confidence level (c.l.) of $98.9\,\%$.
The relation $\log L_{\rm VLBA}=(0.83\pm0.03) \log L_{\gamma} + (5.96\pm0.26)$ can be used as a predictor of $\gamma$-ray flux from radio flux density at 15 GHz.

\item We report a statistically significant positive correlation ($>99\,\%$) 
between $\gamma$-ray luminosity and radio-optical loudness 
($R\propto S_{\rm VLBA}/S_{\rm opt}$) for both quasars and BL~Lacs. 

\item Radio-optical loudness and $\gamma$-optical loudness of the population of quasars is one order of magnitude larger than those of BL Lacs, while no difference is found between $\gamma$-radio loudness of two populations. 

\item We find a correlation between radio luminosity at 15\,GHz and 
$\gamma$-ray-optical loudness ($G_{o} \propto S_{\gamma}/S_{\rm opt}$) for 
quasars, $L_{\rm VLBA} \propto G_{\rm o}^{0.45 \pm 0.12}$ (c.l. $>99.99$\,\%). The relationship between $\gamma$-ray-radio loudness ($G_{r} \propto S_{\gamma}/S_{\rm VLBA}$) and optical nuclear luminosity is marginally significant for BL~Lacs ($L_{\rm opt} \propto  G_{r}^{-0.81\pm 
0.38}$; c.l. $95.8\,\%$) with a large error in the power index.

\item We employ the generic leptonic model for blazars to reproduce the 
observed correlations between $\gamma$-ray emission, optical, and radio 
emission described above. Among all parameters of the jet model, only a 
variation of the accretion power (and the jet power) --- with the magnetic 
field in equipartition with relativistic electrons in the jet --- can 
reasonably well reproduce the observed strong correlations between 
$L_{\gamma}$ and $R$ for quasars and BL Lacs, $L_{\rm opt}$ and $G_{r}$ for BL Lacs, and $L_{\rm VLBA}$ and $G_{o}$ for quasars.
We conclude that the change of the accretion (and jet) power is the main 
parameter responsible for the trends in most of the observed  
 correlations but a combination 
of variations of the accretion power with other jet parameters such as the 
jet viewing angle, Lorentz factor, and the magnetic field is needed to fit all correlations. Independent variations of the latter parameters will open up too large a parameter space, which makes infeasible the consideration of all possible combinations of parameter variations in this work. 
A larger sample of blazars, which will include the faint end of the population, is needed to better constrain the correlations between the bands. Ideally, one would need a simultaneous multi-band monitoring of a larger sample of blazars to test different jet models.

\end{itemize}

\begin{acknowledgements}
We thank the MOJAVE team members for useful discussions. T.G.A. acknowledges support by DFG-SPP project under grant 566960. E.R. acknowledges partial support by the Spanish MICINN through grant AYA2009-13036-C02-02. This work was 
supported by CONACYT research grants 54480 and 151494 (M\'exico), NASA through Fermi Guest Investigator Program award NNX09AT82G and Astrophysics Theory Program Award
NNX10AC79G. The MOJAVE project is supported under National Science Foundation
grant 0807860-AST and NASA-Fermi grant NNX08AV67G. The VLBA is operated by the National Radio Astronomy Observatory a facility of the National Science Foundation operated under cooperative agreement by Associated Universities, Inc. 
\end{acknowledgements}

\bibliographystyle{aa} 

\begin{normalsize}

\begin{longtable}{ccccccccc}

\caption{\label{tab:1} Optical, radio, and $\gamma$-ray parameters of the 76 AGN from the MOJAVE-1 sample. Column (1) is the IAU 1950.0 name of the object; (2) is the optical spectral type: quasars (Q), BL Lacs (BL), and radio galaxies (G); (3) is the redshift; (4) is the logarithm of the total radio flux (radio flux density multiplied by 15~GHz) 
in units of erg s$^{-1}$ cm$^{-2}$; (5) is the logarithm of the nuclear optical flux 
(optical flux density multiplied by 5100\,\AA) 
in units of erg s$^{-1}$ cm$^{-2}$; (6) is the logarithm of the $\gamma$-ray energy flux between 0.1-100~GeV in units of erg s$^{-1}$ cm$^{-2}$; (7)
is the logarithm of the radio luminosity at 15~GHz in units of erg~s$^{-1}$; (8) is the logarithm of the nuclear optical luminosity at 5100\,\AA\ in units of erg~s$^{-1}$; (9) is the logarithm of the $\gamma$-ray luminosity in units of erg~s$^{-1}$. All fluxes are $K$-corrected.} \\

\hline\hline\\ [-2ex]
Name & Sp. type & $z$ & $\log S^{'}_{\rm VLBA}$ & $\log S^{'}_{\rm opt}$ & $\log S^{'}_{\gamma}$ & $\log L_{\rm VLBA}$ & $\log L_{\rm opt}$ & $\log L_{\gamma}$ \\

(1) & (2) &(3) &(4) &(5) &(6) &(7) &(8) &(9) \\

\hline\\ [-2ex]

\endfirsthead
\caption{continued.}\\

\hline\hline\\ [-2ex]

Name & Sp. type  & $z$ & $S^{'}_{\rm VLBA}$ & $S^{'}_{\rm opt}$ & $S^{'}_{\gamma}$ & $\log L_{\rm VLBA}$ & $\log L_{\rm opt}$ & $\log L_{\gamma}$  \\

(1) & (2) &(3) &(4) &(5) &(6) &(7) &(8) &(9) \\

\hline\\ [-2ex]
\endhead
\hline
\endfoot

0059$+$581	& 	 Q 	& 	0.644	& 	 	$-$12.29	& 	$-$12.67	& 	$-$10.27	& 	44.96	& 	44.58	& 	46.97	\\
0106$+$013	& 	 Q 	& 	2.099	& 	 	$-$12.34	& 	$-$12.89	& 	$-$9.82	& 	46.19	& 	45.63	& 	48.70	\\
0109$+$224	& 	 BL	& 	0.265	& 	 	$-$12.81	& 	$-$11.42	& 	$-$10.34	& 	43.52	& 	44.91	& 	45.99	\\
0133$+$476	& 	 Q 	& 	0.859	& 	 	$-$12.12	& 	$-$12.19	& 	$-$9.89	& 	45.44	& 	45.37	& 	47.67	\\
0202$+$149	& 	 Q 	& 	0.405	& 	 	$-$12.51	& 	$-$13.82	& 	$-$11.13	& 	44.25	& 	42.94	& 	45.63	\\
0202$+$319	& 	 Q 	& 	1.466	& 	 	$-$12.46	& 	$-$12.02	& 	$-$10.68	& 	45.68	& 	46.12	& 	47.46	\\
0212$+$735	& 	 Q 	& 	2.367	& 	 	$-$12.38	& 	$-$13.29	& 	$-$10.31	& 	46.27	& 	45.37	& 	48.34	\\
0215$+$015	& 	 Q 	& 	1.715	& 	 	$-$12.84	& 	$-$13.33	& 	$-$10.18	& 	45.46	& 	44.98	& 	48.13	\\
0234$+$285	& 	 Q 	& 	1.207	& 	 	$-$12.20	& 	$-$12.83	& 	$-$10.04	& 	45.72	& 	45.09	& 	47.89	\\
0235$+$164	& 	 BL	& 	0.940	& 	 	$-$12.61	& 	$-$13.20	& 	$-$9.47	& 	45.05	& 	44.46	& 	48.19	\\
0316$+$413	& 	 G 	& 	0.018	& 	 	$-$11.80	& 	$-$10.35	& 	$-$9.76	& 	42.06	& 	43.51	& 	44.09	\\
0336$-$019	& 	 Q 	& 	0.852	& 	 	$-$12.46	& 	$-$12.14	& 	$-$10.70	& 	45.09	& 	45.41	& 	46.84	\\
0403$-$132	& 	 Q 	& 	0.571	& 	 	$-$12.67	& 	$-$12.24	& 	$-$10.88	& 	44.45	& 	44.88	& 	46.24	\\
0415$+$379	& 	 G 	& 	0.049	& 	 	$-$12.05	& 	$-$12.60	& 	$-$10.99	& 	42.69	& 	42.14	& 	43.76	\\
0420$-$014	& 	 Q 	& 	0.914	& 	 	$-$11.79	& 	$-$12.29	& 	$-$10.05	& 	45.84	& 	45.34	& 	47.57	\\
0422$+$004	& 	 BL	& 	0.310	& 	 	$-$12.57	& 	$-$11.76	& 	$-$10.85	& 	43.92	& 	44.73	& 	45.63	\\
0458$-$020	& 	 Q 	& 	2.286	& 	 	$-$12.43	& 	$-$13.01	& 	$-$10.53	& 	46.18	& 	45.61	& 	48.08	\\
0528$+$134	& 	 Q 	& 	2.070	& 	 	$-$11.91	& 	$-$13.25	& 	$-$9.91	& 	46.60	& 	45.26	& 	48.60	\\
0529$+$075	& 	 Q 	& 	1.254	& 	 	$-$12.69	& 	$-$13.04	& 	$-$10.24	& 	45.28	& 	44.93	& 	47.73	\\
0529$+$483	& 	 Q 	& 	1.162	& 	 	$-$12.80	& 	$-$13.30	& 	$-$10.59	& 	45.09	& 	44.59	& 	47.29	\\
0605$-$085	& 	 Q 	& 	0.872	& 	 	$-$12.54	& 	$-$12.73	& 	$-$10.59	& 	45.04	& 	44.85	& 	46.98	\\
0736$+$017	& 	 Q 	& 	0.191	& 	 	$-$12.56	& 	$-$11.95	& 	$-$10.33	& 	43.45	& 	44.06	& 	45.68	\\
0748$+$126	& 	 Q 	& 	0.889	&       	$-$12.32	& 	$-$12.25	& 	$-$10.62	& 	45.28	& 	45.35	& 	46.97	\\
0754$+$100	& 	 BL	& 	0.266	& 	 	$-$12.55	& 	$-$11.32	& 	$-$10.61	& 	43.78	& 	45.01	& 	45.72	\\
0805$-$077	& 	 Q 	& 	1.837	& 	 	$-$12.62	& 	$-$12.55	& 	$-$9.98	& 	45.76	& 	45.83	& 	48.40	\\
0808$+$019	& 	 BL	& 	1.148	& 	 	$-$12.69	& 	$-$12.49	& 	$-$10.78	& 	45.19	& 	45.38	& 	47.09	\\
0814$+$425	& 	 BL	& 	0.245	& 	 	$-$12.71	& 	$-$12.96	& 	$-$10.06	& 	43.55	& 	43.29	& 	46.19	\\
0823$+$033	& 	 BL	& 	0.506	& 	 	$-$12.67	& 	$-$11.47	& 	$-$10.97	& 	44.32	& 	45.52	& 	46.02	\\
0827$+$243	& 	 Q 	& 	0.940	& 	 	$-$12.51	& 	$-$12.09	& 	$-$10.39	& 	45.14	& 	45.56	& 	47.27	\\
0829$+$046	& 	 BL	& 	0.174	& 	 	$-$12.68	& 	$-$11.06	& 	$-$10.54	& 	43.24	& 	44.86	& 	45.38	\\
0836$+$710	& 	 Q 	& 	2.218	& 	 	$-$12.52	& 	$-$12.22	& 	$-$10.12	& 	46.07	& 	46.36	& 	48.46	\\
0838$+$133	& 	 Q 	& 	0.681	& 	 	$-$12.94	& 	$-$12.32	& 	$-$10.92	& 	44.36	& 	44.99	& 	46.39	\\
0851$+$202	& 	 BL	& 	0.306	& 	 	$-$12.20	& 	$-$11.07	& 	$-$10.42	& 	44.28	& 	45.41	& 	46.06	\\
0906$+$015	& 	 Q 	& 	1.024	& 	 	$-$12.38	& 	$-$12.02	& 	$-$10.11	& 	45.37	& 	45.73	& 	47.64	\\
0917$+$624	& 	 Q 	& 	1.446	& 	 	$-$12.86	& 	$-$13.19	& 	$-$10.58	& 	45.26	& 	44.93	& 	47.54	\\
1055$+$018	& 	 Q 	& 	0.890	& 	 	$-$12.09	& 	$-$12.39	& 	$-$10.01	& 	45.51	& 	45.20	& 	47.59	\\
1124$-$186	& 	 Q 	& 	1.048	& 	 	$-$12.36	& 	$-$12.95	& 	$-$10.40	& 	45.41	& 	44.82	& 	47.38	\\
1127$-$145	& 	 Q 	& 	1.184	& 	 	$-$12.28	& 	$-$12.05	& 	$-$10.08	& 	45.62	& 	45.86	& 	47.82	\\
1156$+$295	& 	 Q 	& 	0.729	& 	 	$-$12.30	& 	$-$12.02	& 	$-$10.13	& 	45.09	& 	45.36	& 	47.26	\\
1219$+$044	& 	 Q 	& 	0.965	& 	 	$-$12.84	& 	$-$12.19	& 	$-$10.60	& 	44.84	& 	45.49	& 	47.09	\\
1222$+$216	& 	 Q 	& 	0.432	& 	 	$-$12.86	& 	$-$12.25	& 	$-$10.42	& 	43.96	& 	44.58	& 	46.41	\\
1226$+$023	& 	 Q 	& 	0.158	& 	 	$-$11.35	& 	$-$11.71	& 	$-$9.64	& 	44.48	& 	44.11	& 	46.19	\\
1228$+$126	& 	 G 	& 	0.004	& 	 	$-$12.39	& 	$-$9.78	& 	$-$10.76	& 	40.14	& 	42.76	& 	41.78	\\
1253$-$055	& 	 Q 	& 	0.536	& 	 	$-$11.55	& 	$-$11.65	& 	$-$9.36	& 	45.51	& 	45.40	& 	47.69	\\
1308$+$326	& 	 Q 	& 	0.997	& 	 	$-$12.34	& 	$-$13.09	& 	$-$10.03	& 	45.38	& 	44.63	& 	47.69	\\
1324$+$224	& 	 Q 	& 	1.400	& 	 	$-$13.02	& 	$-$12.51	& 	$-$10.43	& 	45.07	& 	45.58	& 	47.66	\\
1334$-$127	& 	 Q 	& 	0.539	& 	 	$-$11.86	& 	$-$12.59	& 	$-$10.38	& 	45.19	& 	44.47	& 	46.68	\\
1502$+$106	& 	 Q 	& 	1.839	& 	 	$-$12.75	& 	$-$13.08	& 	$-$9.08	& 	45.63	& 	45.30	& 	49.30	\\
1510$-$089	& 	 Q 	& 	0.360	& 	 	$-$12.35	& 	$-$11.61	& 	$-$9.15	& 	44.29	& 	45.03	& 	47.49	\\
1546$+$027	& 	 Q 	& 	0.414	& 	 	$-$12.36	& 	$-$12.27	& 	$-$10.46	& 	44.42	& 	44.51	& 	46.33	\\
1548$+$056	& 	 Q 	& 	1.422	& 	 	$-$12.35	& 	$-$12.58	& 	$-$11.01	& 	45.75	& 	45.53	& 	47.10	\\
1606$+$106	& 	 Q 	& 	1.226	& 	 	$-$12.46	& 	$-$12.52	& 	$-$10.41	& 	45.48	& 	45.42	& 	47.54	\\
1611$+$343	& 	 Q 	& 	1.397	& 	 	$-$12.16	& 	$-$12.41	& 	$-$11.09	& 	45.93	& 	45.67	& 	46.99	\\
1633$+$382	& 	 Q 	& 	1.814	& 	 	$-$12.18	& 	$-$12.66	& 	$-$9.86	& 	46.19	& 	45.71	& 	48.51	\\
1641$+$399	& 	 Q 	& 	0.593	& 	 	$-$11.87	& 	$-$11.66	& 	$-$10.02	& 	45.29	& 	45.50	& 	47.14	\\
1726$+$455	& 	 Q 	& 	0.717	& 	 	$-$12.47	& 	$-$12.66	& 	$-$10.58	& 	44.89	& 	44.70	& 	46.78	\\
1730$-$130	& 	 Q 	& 	0.902	& 	 	$-$11.77	& 	$-$13.55	& 	$-$10.37	& 	45.84	& 	44.06	& 	47.24	\\
1749$+$096	& 	 BL	& 	0.322	& 	 	$-$12.07	& 	$-$12.41	& 	$-$10.10	& 	44.46	& 	44.12	& 	46.42	\\
1803$+$784	& 	 BL	& 	0.680	& 	 	$-$12.42	& 	$-$11.45	& 	$-$10.36	& 	44.88	& 	45.86	& 	46.95	\\
1807$+$698	& 	 BL	& 	0.051	& 	 	$-$12.68	& 	$-$11.84	& 	$-$10.50	& 	42.10	& 	42.94	& 	44.28	\\
1823$+$568	& 	 BL	& 	0.664	& 	 	$-$12.45	& 	$-$12.58	& 	$-$10.41	& 	44.83	& 	44.71	& 	46.87	\\
1828$+$487	& 	 Q 	& 	0.692	& 	 	$-$12.70	& 	$-$12.08	& 	$-$11.12	& 	44.63	& 	45.24	& 	46.21	\\
1849$+$670	& 	 Q 	& 	0.657	& 	 	$-$12.57	& 	$-$11.73	& 	$-$9.78	& 	44.70	& 	45.54	& 	47.49	\\
1958$-$179	& 	 Q 	& 	0.650	& 	 	$-$12.39	& 	$-$12.29	& 	$-$10.29	& 	44.87	& 	44.96	& 	46.96	\\
2131$-$021	& 	 BL	& 	1.285	& 	 	$-$12.47	& 	$-$12.76	& 	$-$11.08	& 	45.52	& 	45.24	& 	46.91	\\
2145$+$067	& 	 Q 	& 	0.990	& 	 	$-$11.80	& 	$-$11.91	& 	$-$10.81	& 	45.91	& 	45.80	& 	46.90	\\
2155$-$152	& 	 Q 	& 	0.672	& 	 	$-$12.48	& 	$-$12.25	& 	$-$10.89	& 	44.81	& 	45.04	& 	46.41	\\
2200$+$420	& 	 BL	& 	0.069	& 	 	$-$12.06	& 	$-$11.29	& 	$-$10.09	& 	42.99	& 	43.76	& 	44.96	\\
2201$+$171	& 	 Q 	& 	1.076	& 	 	$-$12.51	& 	$-$12.98	& 	$-$10.16	& 	45.29	& 	44.82	& 	47.64	\\
2209$+$236	& 	 Q 	& 	1.125	& 	 	$-$12.61	& 	$-$13.09	& 	$-$10.88	& 	45.24	& 	44.76	& 	46.97	\\
2223$-$052	& 	 Q 	& 	1.404	& 	 	$-$12.13	& 	$-$12.52	& 	$-$10.31	& 	45.96	& 	45.57	& 	47.78	\\
2227$-$088	& 	 Q 	& 	1.560	& 	 	$-$12.48	& 	$-$12.46	& 	$-$9.83	& 	45.72	& 	45.75	& 	48.38	\\
2230$+$114	& 	 Q 	& 	1.037	& 	 	$-$12.15	& 	$-$12.13	& 	$-$10.04	& 	45.61	& 	45.64	& 	47.72	\\
2251$+$158	& 	 Q 	& 	0.859	& 	 	$-$11.81	& 	$-$11.88	& 	$-$9.09	& 	45.75	& 	45.68	& 	48.47	\\
2331$+$073	& 	 Q 	& 	0.401	& 	 	$-$12.72	& 	$-$11.73	& 	$-$10.82	& 	44.03	& 	45.02	& 	45.93	\\
2345$-$167	& 	 Q 	& 	0.576	& 	 	$-$12.41	& 	$-$12.53	& 	$-$10.63	& 	44.72	& 	44.60	& 	46.50	\\

\end{longtable}

\end{normalsize}

\end{document}